# Entangled and correlated photon mixed strategy for social decision making

Shion Maeda[1,*], Nicolas Chauvet[1,2], Hayato Saigo[3], Hirokazu Hori[4],

Guillaume Bachelier[5], Serge Huant[5], and Makoto Naruse[1,2,**]

[1] Department of Mathematical Engineering and Information Physics, School of Engineering, The University of Tokyo, 7-3-1 Hongo, Bunkyo-ku, Tokyo 113-8656, Japan

[2] Department of Information Physics and Computing, Graduate School of Information Science and Technology, The University of Tokyo, 7-3-1 Hongo, Bunkyo-ku, Tokyo 113-8656, Japan

[3] Nagahama Institute of Bio-Science and Technology, 1266 Tamura, Nagahama, Shiga 526-0829, Japan

[4] Interdisciplinary Graduate School, University of Yamanashi, Takeda, Kofu, Yamanashi 400-8510, Japan

[5] Univ. Grenoble Alpes, CNRS, Institut Néel, 38000 Grenoble, France

* maeda-shion4141@g.ecc.u-tokyo.ac.jp  ** makoto_naruse@ipc.i.u-tokyo.ac.jp

**Abstract**

Collective decision making is important for maximizing total benefits while preserving equality among individuals in the competitive multi-armed bandit (CMAB) problem, wherein multiple players try to gain higher rewards from multiple slot machines. The CMAB problem represents an essential aspect of applications such as resource management in social infrastructure. In a previous study, we theoretically and experimentally demonstrated that entangled photons can physically resolve the difficulty of the CMAB problem. This decision-making strategy completely avoids decision conflicts while ensuring equality. However, decision conflicts can sometimes be beneficial if they yield greater rewards than non-conflicting decisions, indicating that greedy actions may provide positive effects depending on the given environment. In this study, we demonstrate a mixed strategy of entangled- and correlated-photon-based decision-making so that total rewards can be enhanced when compared to the entangled-photon-only decision strategy. We show that an optimal mixture of entangled- and correlated-photon-based strategies exists depending on the dynamics of the reward environment as well as the difficulty of the given problem. This study paves the way for utilizing both quantum and classical aspects of photons in a mixed manner for decision making and provides yet another example of the supremacy of mixed strategies known in game theory, especially in evolutionary game theory.

**Introduction**

Optics and photonics are expected to play key roles in accommodating the massive requirements of future intelligent information systems [1]. Recent developments in the field include photonic reservoir computing for time series predictions [2–4], on-chip lightwave circuits for photonic learning functions [5,6], fibre optic neuromorphic systems [7,8], among others. Decision making is another important research topic in which decisions have to be made autonomously in dynamically changing uncertain environments [9,10]. Furthermore, collective decision making involving multiple players becomes a critical issue in the management of social utilities [11]. Recently, photonic approaches to decision-making problems have been intensively studied using single photons [12], chaotic lasers [13], and entangled photons [14].

The multi-armed bandit (MAB) problem describes some of the fundamental issues associated with decision making. The objective of the MAB problem is to maximize the player rewards from multiple slot machines with an initially unknown hit probability. Here, spending a lot of time gathering new information can be costly, while hasty decisions lead to missing out on good choices. This issue is known as the exploration-exploitation dilemma [15]. The physical attributes of photons have been successfully utilized to solve such MAB problems [16]. The MAB problem becomes even more complicated when multiple players come into play. As individual players seek to maximize their rewards, they will choose the best machine, which leads to a decision conflict. Many relevant issues in real-life situations, ranging from congestion in information networks and traffic jams on roads to hoarding of goods are caused by many players choosing the same decision [9,10], suggesting the importance of collective decision making. Such a problem that deals with multiple players and slot machines is called the competitive multi-armed bandit (CMAB) problem [14].

In previous work concerning the two-player, two-armed bandit problem, we theoretically and experimentally demonstrated collective decision making by using polarization-entangled photon pairs so that decision conflicts are avoided and maximum total reward is accomplished while ensuring equality [14]. More recently, we have theoretically derived optimal quantum states that provide the maximum total reward, while preserving equality, with respect to more than three players on two-armed bandit problems [17]. We also showed that classical photons, such as single photons and correlated photon pairs, cannot resolve decision conflicts [14]. In these studies, the reward dispensed from a slot machine is constant at a time. Hence, in the event of a decision conflict, the individual player's reward is reduced by the number of overlaps, resulting in a reduction in the total reward.

However, depending on the given environmental conditions, decision conflicts can also provide a greater total reward. For example, if the individual reward is *not* reduced from a particular slot machine even in the case of conflicts, choosing the same slot machine (decision conflict) yields a higher total reward than choosing different slot machines (non-conflicting decision). Indeed, we can observe similar real-life scenarios, for example, in the form of enhanced services or resource availability, such as computing power and reduced sales prices which are offered during a limited amount of time.

Similarly, the notion of critical mass that has to be reached for an activity to be sustainable reflects the non-decreasing rewards with conflicting decisions.

In this study, to accommodate the aforementioned changes in the environmental conditions and maximize total rewards, we propose and demonstrate a mixed strategy of utilizing entangled photons and classical photons (specifically, polarization-entangled photon pairs and polarization-correlated photon pairs) to find the optimal solution of 2-player, 2-armed bandit problems. While utilizing entangled photons, which guarantee non-conflicted and fully equal decisions, each player accumulates information about the reward environment. When recognizing that the conflicted choice provides greater rewards, we utilize correlated photons to fully exploit the reward from the environment. We show that an optimal mixture of entangled and correlated photons exists depending on the dynamics of the reward environment as well as the difficulty of finding the higher reward probability machine. Although the following discussion is restricted to 2-player, 2-choice problems, the present study captures the essential aspects of entangled and classical-photon mixed strategies that can be extended for solving more generalized problems.

## Results

### System architecture

We consider two players (Players 1 and 2), each of whom chooses one of two slot machines (Machines A and B) with the intention of maximizing the total reward or the summation of the reward of each player. The reward probabilities of Machines A and B are denoted as $P_A$ and $P_B$, respectively. Although the present study examines the properties of entangled and classical photon states theoretically and numerically, it assumes technologically feasible experimental optical systems to generate photon pairs by spontaneous parametric down conversion (SPDC), as schematically represented in Fig. 1a, which is essentially the same as the experimental setup proposed in our previous study [14]. The photon pair generation is based on a standard Sagnac loop architecture [18] to induce SPDC. The signal and idler photon correspond to the decisions of Players 1 and 2, respectively. The signal photon goes through a half-wave plate ($HW_1$) followed by a polarization beam splitter PBS ($PBS_1$). If the photon is detected by the photodetector corresponding to the horizontally polarized light (PD1), the decision of Player 1 is to choose Machine A, whereas if the photon is detected by the photodetector corresponding to the vertically polarized light (PD2), then the decision of Player 1 is to choose Machine B. Similarly, the decision of player 2 is determined by the detection of the idler photon by PD3 or PD4, which corresponds to the decisions of selecting Machines A and B, respectively.

We introduce several notations to describe the system. The input photon state for the decision of Player $i$ ($i=1,2$) is denoted as $|\theta_i\rangle$, where $\theta_i$ is the linear polarization angle. The roles of $HW_i$ and $PBS_i$ are given by

$$HW_i|\theta_i\rangle = |2\theta_{HW_i} - \theta_i\rangle \tag{1}$$

and

$$PBS_i|2\theta_{HW_i} - \theta_i\rangle = \cos(2\theta_{HW_i} - \theta_i)|H_i\rangle + \sin(2\theta_{HW_i} - \theta_i)|V_i\rangle, \tag{2}$$

where $|H_i\rangle$ and $|V_i\rangle$ indicate photon states with horizontal and vertical polarization propagating in orthogonal directions beyond PBS$_i$ [19]. One strategy for realizing collective decision making is to link the decisions of Players 1 and 2 by introducing correlations among the decisions at the level of photon states. Here, we consider polarization-orthogonal photon pairs denoted by, $|\theta_1, \theta_2\rangle$ where

$$\theta_2 = \theta_1 + \pi/2, \tag{3}$$

as input photon states to the two players. In practice, we can fix $\theta_i$ ($i = 1,2$) by controlling the polarizers and half/quarter waveplates in the path of the excitation laser (respectively denoted by P, HW$_E$, and QW$_E$ in Fig. 1a). Let us set $\theta_1 = 0$ and $\theta_2 = \pi/2$, for the sake of simplicity. The probability of observing photons at PD1 and PD3 (meaning that both players chose Machine A) and at PD2 and PD4 (both players select Machine B) are represented by the following equations.

$$P_C(A,A) = \cos^2\left(2\theta_{HW_1}\right)\cos^2\left(2\theta_{HW_2} - \frac{\pi}{2}\right) \tag{4}$$

and

$$P_C(B,B) = \sin^2\left(2\theta_{HW_1}\right)\sin^2\left(2\theta_{HW_2} - \frac{\pi}{2}\right). \tag{5}$$

By letting $\theta_{HW_1} = 0$ and $\theta_{HW_2} = \pi/2$ or their $N\pi$ angle-shifted equivalents (where $N$ is an integer) in Eq. (4), $P(A,A)$ becomes unity, indicating that both players always chose Machine A, which is schematically illustrated in Fig. 1b. Similarly, $P(B,B)$ becomes unity when $\theta_{HW_1} = \pi/2$ and $\theta_{HW_2} = 0$ or their $N\pi$ angle-shifted equivalents in Eq. (5). The probability of observing photons at PD1 and PD4 is given by

$$P_C(A,B) = \cos^2\left(2\theta_{HW_1}\right)\sin^2\left(2\theta_{HW_2} - \frac{\pi}{2}\right) \tag{6}$$

which becomes unity when $\theta_{HW_1} = 0$ and $\theta_{HW_2} = 0$ or their $N\pi$ phase-shifted equivalent. $P_C(A,B)=1$ implies that Player 1 always chooses Machine A while Player 2 always selects Machine B. There is indeed no decision conflict in this case, but equality is severely deteriorated. More details can be found in Ref. [14].

To overcome this issue, we utilize a coherent superposition of states corresponding to the entangled states. Here, we consider the maximally entangled singlet photon state given by

$$\frac{1}{\sqrt{2}}\left(|\theta_1,\theta_2\rangle - |\theta_2,\theta_1\rangle\right), \tag{7}$$

where $\theta_1$ and $\theta_2$ are orthogonal to each other, as specified in Eq. (3). The maximally entangled photons are usually represented in the form $\frac{1}{\sqrt{2}}(|HV\rangle - |VH\rangle)$. A different notation is used in Eq. (7) to maintain consistent notations with the aforementioned polarization-correlated photons ($|\theta_1,\theta_2\rangle$) and to clearly present the role of half-wave plates in the following discussion. Considering the probability amplitude originating from the second term in Eq. (7), the probabilities of the two players' decisions are given by [14]:

$$P_E(A,A) = P_E(B,B) = \frac{1}{2}\sin^2\left[2\left(\theta_{HW_1} - \theta_{HW_2}\right)\right]. \tag{8}$$

$$P_E(A,B) = P_E(B,A) = \frac{1}{2}\cos^2\left[2\left(\theta_{HW_1} - \theta_{HW_2}\right)\right], \tag{9}$$

which means that if $\theta_{HW_1} = \theta_{HW_2}$ is satisfied, the non-conflict probability is always unity and equality is ensured by Eq. (9). This also means that the conflict probability is always zero in Eq. (8) regardless of the values of $\theta_i$ and $\theta_{HW_i}$. Such collective decision-making is schematically illustrated in Fig. 1b.

If $\theta_{HW_1}$ and $\theta_{HW_2}$ are orthogonally arranged with each other, for example, $\theta_{HW_1} = \theta_{HW_2} + \pi/2$, the relationship is completely reversed: decision conflict is always induced with equal probability at Machine A and Machine B. Although such an orthogonally arranged configuration is another interesting aspect of entangled-photon states given by Eq. (7), it is not exploited in the following discussion for simplicity.

When two players make the same decision, the reward for each player is usually divided into two halves, as schematically illustrated in Fig. 2a. Therefore, from the viewpoint of maximizing the total reward, when a player chooses the best slot machine, the other player should select the other machine. Hence, entangled-photon-based decision making theoretically provides the maximum total reward [14]. However, as discussed in the introduction, decision conflicts could yield greater total reward depending on the reward environment conditions. Here, we define the notion of a *happy hour*. During the happy hour, one of the two slot machines dispenses a reward of unity per play to all players who select that machine even when the decisions are conflicted, as schematically shown in Fig. 2b. In the present study, we assume that the higher-reward-probability machine occasionally provides a happy hour. This means that a player gets one coin when he wins even if the decision is in conflict with the decision of the other player during the happy hour. At the same time, it should be emphasized that the reward per play is unity during non-happy hours. That is, a player can get one coin when he wins if the decision is not conflicted. Therefore, the player *cannot* detect the occurrence of a happy hour simply by observing the amount of reward per play. On the other hand, the player can immediately realize the end of the happy hour because the dispensed reward decreases to one-half due to decision conflict.

**Mixed strategy**

The aim of the present study is to statistically *mix* entangled-photon-based decision-making and correlated-photon-based decision making. While the entangled photons provide non-conflict decisions, the players can accumulate information about the slot machines. Assume that Machine $i$ is selected $N_i$ times, and the number of wins is $L_i$. Based on maximum likelihood estimation, the estimated reward probability of Machine $i$ is given by $\hat{P}_i = L_i/N_i$ ($i$ = A, B). Here, we consider that the machine that gives the maximum $\hat{P}_i$ would be the best machine. This machine is denoted as Machine $m$. The source photon states are switched (by HW$_E$ and QW$_E$ in Fig. 1a) so that they provide correlated photons. Although here we tune a common photon pair source for both types of states, this could be done by switching from one distinct photon pair source to another, without any difference for the results. At the same time, the half-wave plates of Players 1 and 2 (denoted, respectively, by HW$_1$ and HW$_2$ in Fig. 1a) are configured in such a way that Machine $m$ is chosen based on Eqs. (4) and (5), i.e., conflicting decision-making is intentionally induced. If the amount of reward is unity in such an intentionally induced conflicted decision, we can deduce that Machine $m$ is indeed being operated at a happy hour. In addition, once Machine $m$ returns to the non-happy hour operation, the player can immediately detect the end of the happy hour since the dispensed reward becomes one-half due to decision conflict. Such a mixed strategy of entangled- and correlated-photons is summarized in Algorithm 1 and Fig. 2c.

**Algorithm 1**: Entangled and correlated-photon mixed strategy

1. [Entangled strategy] Play slot machines based on entangled-photon-based decision making while accumulating knowledge about the reward probability of the slot machines. Repeat this strategy during *SI* steps. *SI* refers to search interval. Determine the highest reward probability machine by $m = \arg\max_i(\hat{P}_i)$.

2. [Correlated strategy] Play slot machines based on correlated-photon-based decision-making. Here, the half-wave plates of Players 1 and 2 are configured so that both players select Machine *m*. Repeat this strategy during *CP* steps. *CP* refers to check span.

3. If the dispensed reward does not become unity, go back to the entangled-photon decision making (Step 1).

   If the dispensed reward is unity (i. e., Machine *m* is operated in a happy hour), the correlated-photon strategy is maintained. When the dispensed reward becomes half, go back to the entangled-photon decision making (Step 1).

## Discussion

The present study is stimulated by the notion of an evolutionary stable strategy (ESS) known in evolutionary game theory [20]. Let us roughly describe the Hawk-Dove game to convey a fundamental concept of ESS. Detailed discussions of this strategy are available in literature, such as Ref. [20].

Players who choose the Hawk strategy always take adversarial actions when confronted by their opponents. By doing so, they can gain a lot of rewards if they win the battle. However, they could also suffer from huge damage if they lose. Conversely, players who choose the Dove strategy always avoid battles when they face their enemies. Hence, there is no gain (because they avoid battles), but there is also no risk of loss. In evolutionary game theory, there exists an optimal mixture of Hawk and Dove strategies that maximizes the expected rewards, and this mixed strategy can be superior to both pure Hawk and pure Dove strategies depending on the environment. The optimal mixture depends on the gains and losses in the battle.

We observe similarities between this concept in evolutionary game theory and the present study of the CMAB problem. The Dove strategy is similar to the entangled-photon strategy which attempts to secure the achievable total reward, while the Hawk strategy is like the correlated-photon strategy which seeks greater reward at a certain degree of risk. The difference lies in the method by which the optimal mixture is derived.

In the following numerical analysis, 1500 consecutive slot machine plays are conducted for the 2-player, 2-armed CMAB problem. For the sake of simplicity, we assume a fixed reward probability throughout the 1500 plays for $P_A$ and $P_B$. The total reward, which is the summation of the rewards gained by Player 1 and Player 2, is calculated by averaging over 1,000 repetitions of such 1500 consecutive plays. See the *Methods* section for the details. In addition, we assume that $P_A$ is greater than $P_B$ while the condition of $P_A + P_B = 1$ holds. Therefore, if there are no happy hours, the expected maximum total reward is 1500 by the entangled-photon decision strategy. This is because the entangled photons ensure the absence of conflict, meaning that the two machines are always selected. The dashed blue line in Fig. 3a shows the calculated total reward.

Now, we examine the impact of the occurrence of happy hours. We assume that happy and non-happy hours periodically interchange in every $T$ steps, with $T$ being an integer. Let us first focus on the case when $T$ is equal to 50. The red curve shown in Fig. 3a represents the total reward as a function of the search interval when $(P_A, P_B) = (0.6, 0.4)$. We can observe that the total reward is greater than that of the entangled-photon-only strategy if $SI$ is between 6 and 30, while the maximum total reward is realized when $SI = 14$. Hence, $SI = 14$ provides the optimal mixed strategy for this particular reward environment. A search interval that is too short indicates excessive greedy actions ($SI < 6$), whereas an excessively long search interval indicates missing a large reward during the happy hours ($SI > 30$).

The green, magenta, and brown curves in Fig. 3a show the total reward when $(P_A, P_B)$ is equal to (0.7, 0.3), (0.8, 0.2), and (0.9, 0.1), respectively. The optimal search interval that yields the maximum total reward decreases as $P_A$ becomes larger. This is because the rewards gained during happy hours dramatically increase as $P_A$ increases. When $P_A = 0.9$, the total reward is almost 2050, which is nearly a 40% increase compared with the entangled-photon-only strategy. In addition, when $P_A$ is greater than 0.7, the total reward is greater than the entangled-photon-only strategy even with extremely small as well as extremely long search intervals, indicating that the gain accomplished during a happy hour surely pays off the cost of conducting correlated-photon-based greedy actions.

The optimal mixture, however, also depends on the frequency of happy hours. The curves in Fig. 3b examine the total rewards when the interval of the happy hour is increased from 10 to 100 steps. First, we consider the case when $(P_A, P_B) = (0.6, 0.4)$ denoted by Fig. 3b-iv. As the happy-hour interval decreases, the maximum total reward decreases, indicating that the mixed strategy cannot adapt to rapid environmental changes. Nevertheless, we also observe that the search interval that yields the maximum total reward decreases as the happy-hour interval decreases, meaning that frequent usage of a correlated-photon-based decision-making strategy provides a greater total reward. The same tendency is observed in different reward probability settings $(P_A, P_B)$ given by (0.9, 0.1), (0.8, 0.2), and (0.7, 0.3), which are summarized in Figs. 3b-i, ii, and iii, respectively. Furthermore, it is interesting to observe the oscillatory behaviour of the total rewards as a function of the search interval. This is clearly due to interdependence between the environmental switching and the strategy switching as the period of oscillation is about twice the period of environmental switching and does not depend on the reward probabilities.

For the sake of obtaining higher rewards regardless of the given environmental change dynamics and to accommodate any uncertainty in the given environment, we discuss the optimality of the search interval. The curves in Fig. 3c represent the normalized total reward, $(R - R_{MIN})/(R_{max} - R_{MIN})$, where $R$ is the total reward for a search interval. $R_{MIN}$ and $R_{MAX}$ indicate the minimum and maximum total rewards in the range of the search interval under study ($1 \leq SI \leq 50$). The black curves represent the average total reward over different happy-hour intervals. The search interval that maximizes the normalized total rewards is summarized in Fig. 4a. The horizontal axis represents the difficulty of finding a higher reward probability machine, which is defined as $1 - (P_A - P_B)$. The reward probability of the combination $(P_A, P_B) = (0.9, 0.1)$ corresponds to a difficulty of 0.2, whereas $(P_A, P_B) = (0.5, 0.5)$ corresponds to a difficulty of unity, which means that the slot machines are identical. One remark here is that Machine A provides happy hour in the case of $P_A = P_B$ while Machine B does not provide happy hour. The optimal search interval monotonically increases as the difficulty of finding a better machine increases. Furthermore, if the difficulty is less than 0.8 (the reward probability differences are greater than 0.2), the search interval of approximately 5 to 10 is close to the optimal search interval. This is also confirmed by Fig. 3b, suggesting that such a search interval can accommodate the uncertainty of reward environments in terms of both the reward probability values as well as the dynamic change of happy hour occurrences.

The check span is another critical figure that influences the optimal mixture of entangled and correlated-photon-based decision-making strategy. Figure 4b examines the total rewards as a function of the check span assuming $(P_A, P_B) = (0.7, 0.3)$ while considering the happy hour interchange intervals ranging from 10 to 100. The maximum total reward is obtained by a check span of 2. This result indicates that too short an CP ($CP < 2$) may miss the detection of a happy hour because of the probabilistic attributes of the slot machine, whereas excessive CP ($CP > 2$) also leads to excessive loss. The analysis in Fig. 3 was conducted based on an CP value of 2.

In this work, we focussed on the 2-player, 2-machine CMAB problem to highlight the central concept and principle of the entangled and correlated-photon mixed strategy. The extension of the proposed method to a general case, $N$-player, $M$-machine CMAB is an important future study. Indeed, Chauvet *et al.* have already demonstrated optimal entangled photon states for three, four, and five players for the 2-armed bandit problem [17]. Scalability analysis is also critical from the viewpoint of practical applications such as resource management in information and communication infrastructure [10]. Also, the mixed strategy studied herein allows the players to immediately change the photon states from correlated to entangled photons when they detect the end of happy hour in Step 3 of Algorithm 1. Such controllability or accessibility of the photon source by the player can be generalized through using time delay for example, which is an interesting future topic.

The reward probability estimation also needs to be studied. In Step 1 of Algorithm 1, the expected reward probabilities of Machine A and B were evaluated as $\hat{P}_A = L_A/SI$ and $\hat{P}_B = L_B/SI$, respectively, after *SI*-steps slot machine play. Here, $L_A$ and $L_B$ denote the number of wins by playing Machines A and B, respectively. It is remarkable that the denominators of $\hat{P}_A$ and $\hat{P}_B$ are both *SIs* because the decisions were based on entangled photons, i.e., Machine A and Machine B were chosen exactly the same number of times. However, it should also be noted that in the present study, the information integration is assumed from both players. In a future study, we will tackle the case where the reward probability estimation is completely independently conducted by each player. We would also presume, however, that the impact of this independent estimation may be minor, particularly when *SI* is moderately large owing to the equality secured by entangled photons. It should also be noted that order recognition is required for solving general $N$-player, $M$-machine CMAB problems, especially when $M > N$. Hence, finding the highest reward probability machine alone is not sufficient, and a novel strategy should be developed for order recognition. To achieve this, several approaches, such as using confidence interval [21] and Schubert calculus [22], could be integrated with the present study in the future.

**Conclusion**

We theoretically and numerically demonstrated an entangled and correlated-photon-based mixed decision-making strategy to obtain enhanced total rewards in dynamically changing reward environments. Entangled-photon-based decision-making completely avoids conflict and secures equal opportunities for all players. However, conflict avoidance does not necessarily maximize total rewards in reward environments where greedy actions are beneficial, even socially. By introducing the notion of happy hours in competitive multi-armed bandit problems, the cases in which conflicts are beneficial are systematically examined. We demonstrated an optimal mixture of entangled and correlated-photon strategies in terms of adequate switching intervals between these two strategies. The present study is relevant to evolutionary stable states known in evolutionary game theory, where the optimally mixed strategy provides greater expected rewards than other mixed and pure strategies in biological species. We observe similarities between the proposed method and evolutionary game theory in terms of the mixture in strategies themselves as well as the dependence on the given environment. This study paves the way for utilizing both quantum and classical aspects of photons in a mixed manner as well as demonstrating yet again, the supremacy of mixed strategies.

## METHODS

**Numerical analysis.** The numerical analysis of the present study was conducted on a personal computer (MacBook Pro, Intel Core i5, 1867 MHz, 8 GB RAM, macOS Catalina, MATLAB R2019a). Here, we describe the details of the statistical evaluation of the entangled- and correlated-photon mixed strategy for CMAB problems. For emulating the entangled photons and the two slot machines, we utilized uniformly distributed pseudorandom numbers generated by Mersenne twister. When the interval of happy and non-happy hours is specified by the interval $T$, the initial starting time of the happy hour was randomly determined by a uniformly distributed random natural number between 1 and $2 \times T$, for each repetition. The results shown in Figs. 3 and 4 were obtained by averaging over 1,000 such randomly arranged repetitions.

## ACKNOWLEDGEMENTS

This work was supported in part by the CREST project (JPMJCR17N2) funded by the Japan Science and Technology Agency, the Core-to-Core Program A. Advanced Research Networks and Grants-in-Aid for Scientific Research (A) (JP20H00233) funded by the Japan Society for the Promotion of Science.

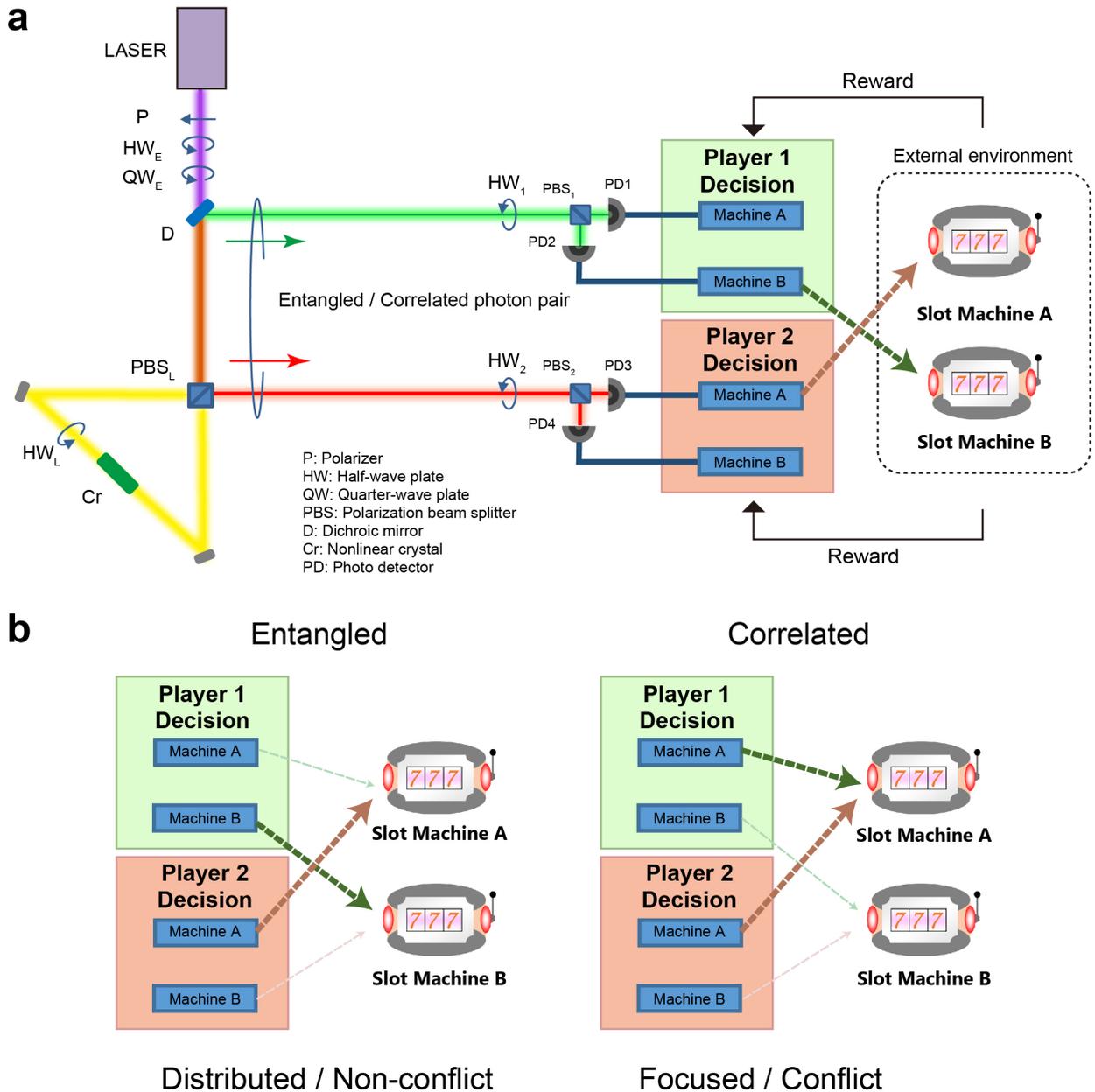

**Fig. 1 | System architecture for entangled and correlated-photon mixed strategy for collective decision making.** (a) Schematic of the optical system configuration consisting of photon pair (either entangled or correlated) generation and photon detection systems that directly provide the decisions of the two players (Player 1 and 2) to select either of the two slot machines (Machine A or Machine B) in the external environment. (b) With entangled photons, the decisions of the players are never in conflict. With correlated photons, both players take the same decision by properly choosing the half waveplate angle (see text for details). The elements of Fig. 1a are adapted from Chauvet et al., Sci. Rep. 9, 12229 (2019). Copyright 2019 Author(s), licensed under a Creative Commons Attribution 4.0 License.

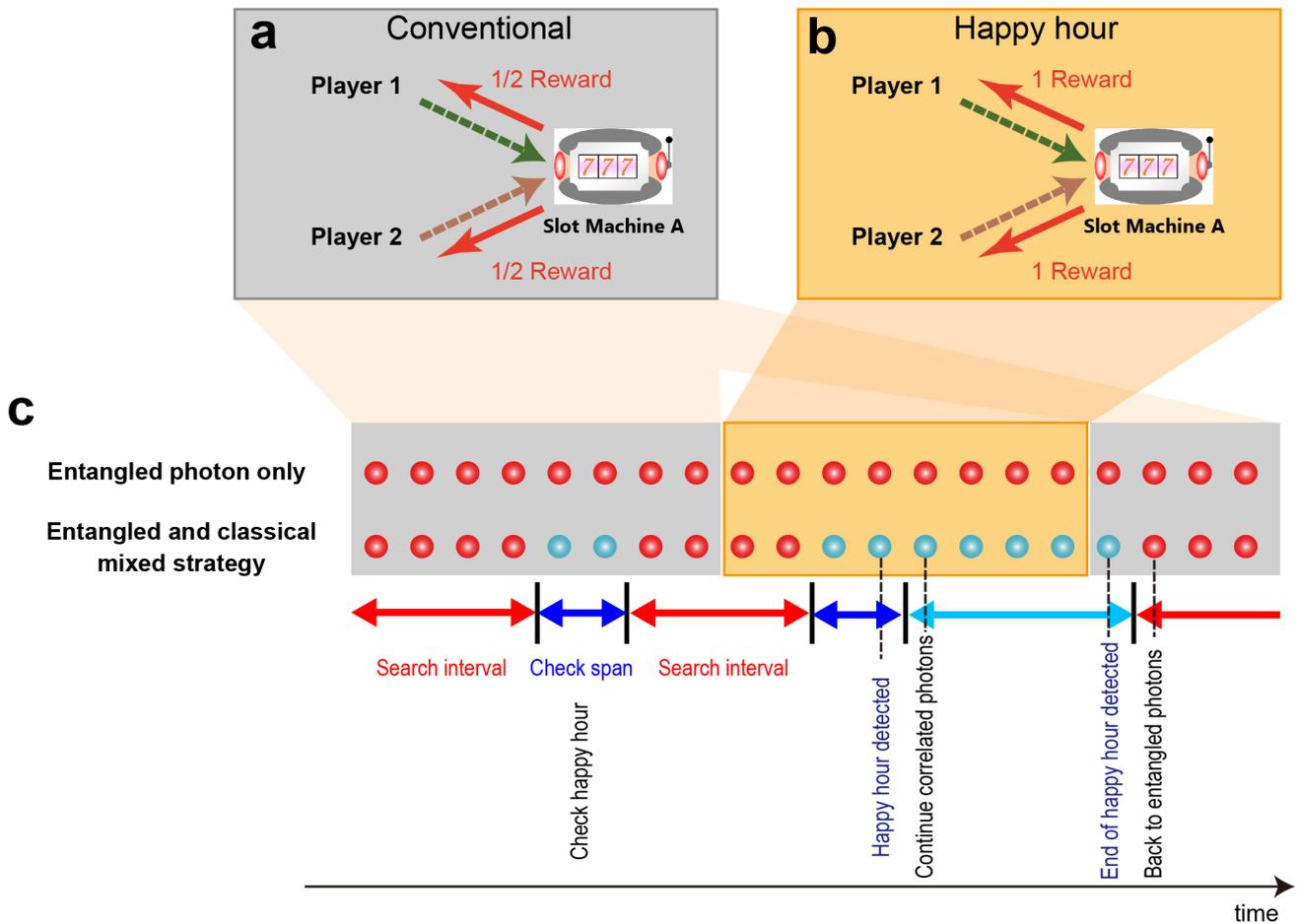

**Fig. 2 | Reward environments and collective decision-making strategies.** (a) Usually, the reward gained by a single player is half a unit reward, if the decision is conflicted. (b) During the happy hour, the higher-reward probability machine yields unit reward to both the players, even if the decision is conflicted. (c) In the mixed strategy, while entangled photons are used most of the time, correlated photons are occasionally used (denoted by search interval) during certain duration (check span) to check if greedy action is more beneficial.

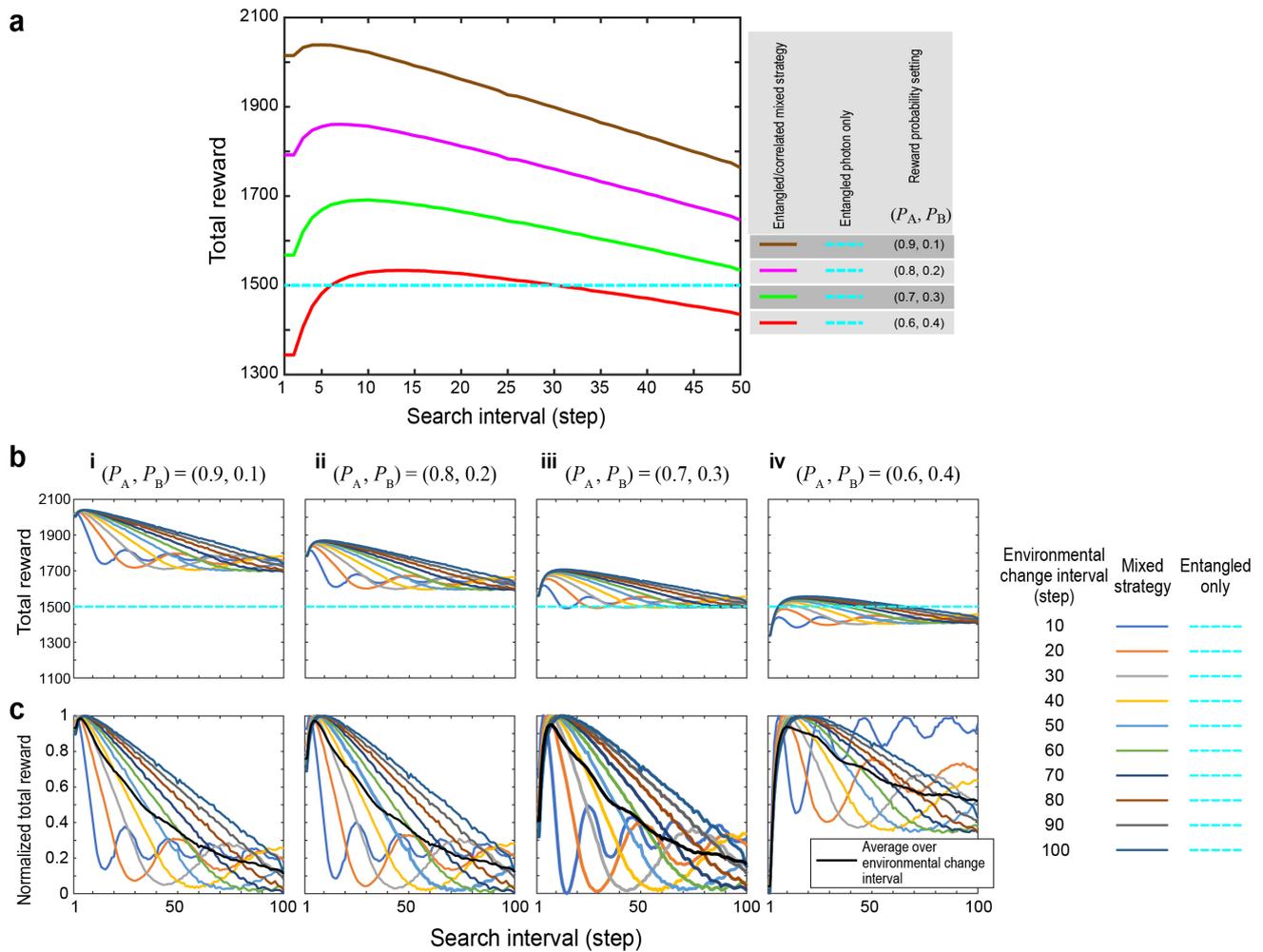

**Fig. 3 | Demonstration of the proposed mixed strategy.** (a) The dashed cyan line shows the total reward gained by using the entangled-photon-only strategy. With the mixed strategy, the total reward can be greater than that of the entangled-photon-only decision-making strategy. Here, the happy hour and non-happy hour are periodically switched every 50 steps. (b) The dependence of the total reward on the environmental change interval, and (c) their normalized representation to examine optimality of the mixture of entangled and correlated photons for a variety of environmental conditions.

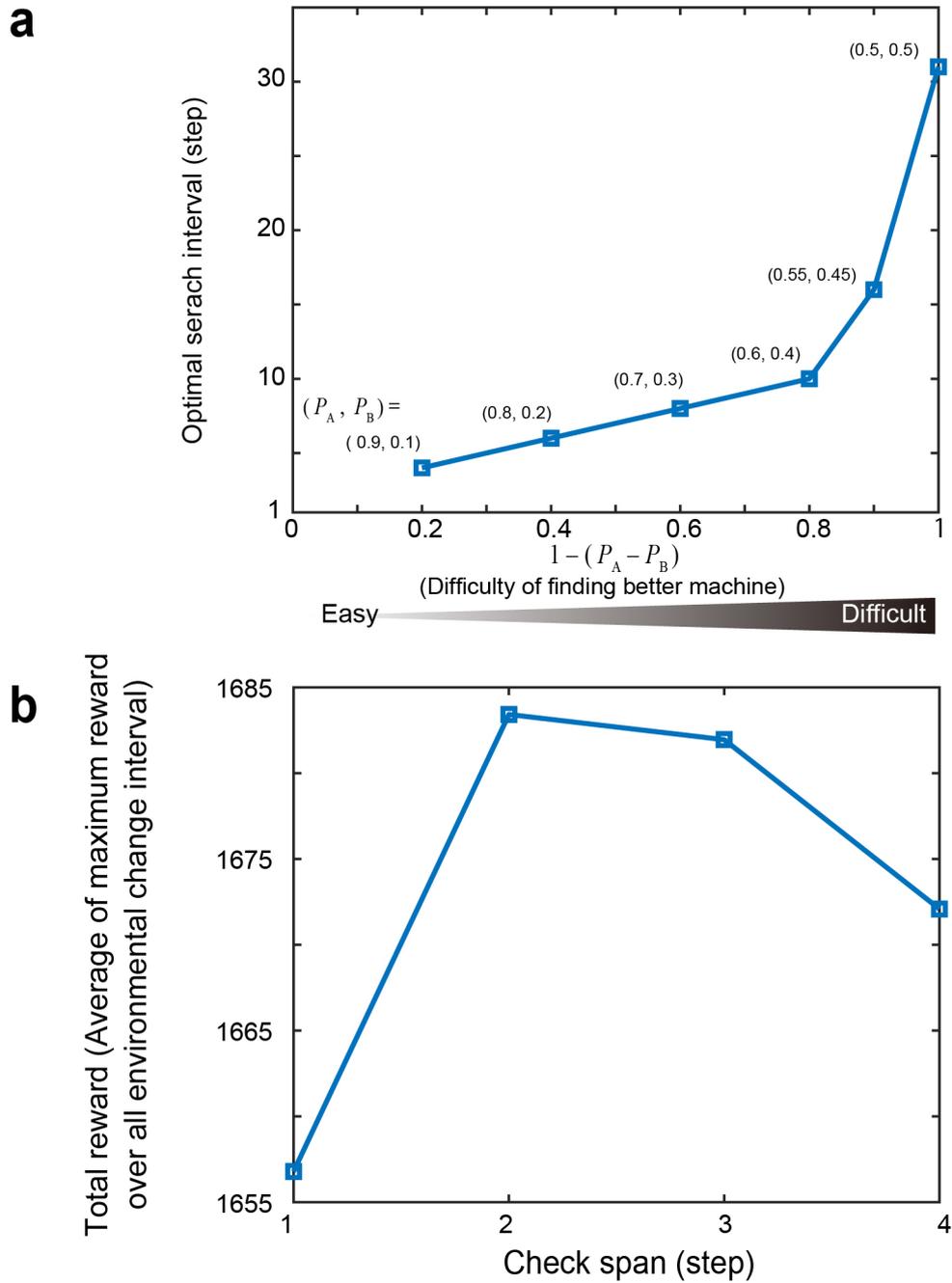

**Fig. 4 | Optimal mixture of entangled- and correlated-photon-based decision making.** (a) The optimal search interval monotonically increases as the difficulty of finding better slot machine increases. Also, if the difficulty is less than 0.8, a search interval of approximately 5 to 10 yields nearly optimal total rewards (See Fig. 3b). (b) The check span is another important parameter for the mixed strategy. Check span of 2 results in the maximum total rewards.